\documentclass[a4,12pt]{article}
\usepackage{graphics}
\usepackage[percent]{overpic}

\usepackage{overcite}

\makeatletter
\renewcommand{\@biblabel}[1]{#1 }
\makeatother

\begin{document}

\title{Nanostructure for Hybrid Plasmonic-Photonic Crystal Formed on Gel-Immobilized Colloidal Crystal Observed by AFM after Drying}

\author{Sho Kawakami$^{1)}$, Atsushi Mori$^{2)*}$, Ken Nagashima$^{3)}$, \\
Shuichi Hashimoto$^{2)}$, Masanobu Haraguchi$^{2)}$ \\[1em]
1) Department of Optical System Engineering, Tokushima University, \\
2-1 Minamijosanjima, Tokushima, 770-8506, Japan \\
2) Institute of Technology and Science, Tokushima University,\\
 2-1 Minamijosanjima, Tokushima, 770-8506, Japan \\
3) Institute of Low Temperature Science, Hokkaido University, \\
Kita-19, Nishi-8, Kita-ku, Sapporo, 060-0819, Japan}

\maketitle

\noindent
Received: 14 August, 2015; revised:\\
E-mail: atsushimori@tokushima-u.ac.jp
\newpage

\begin{abstract}
Aiming at fabrication of hybrid plasmonic-photonic crystals, gel-immobilized colloidal crystals made of a polystyrene colloidal suspension and an $N$-(Hydroxy methyl)acrylamid-based gel were immersed into an aqueous dispersion of gold nanoparticles.
Atomic force microscope (AFM) observations have been performed for the gel-immobilized colloidal crystals with gold nanoparticles deposited on their surfaces.
In the present study, the diameter of a colloidal sphere was c.a.~190 nm.
The diameter of a gold nanoparticle was the same as in a preliminary study, c.a.~40 nm.
Various immersion times up to two hours were tested.
Surface of a sample of 2 hr immersion has been observed.
Prior to the AFM observation, the sample was dried in a desiccator for 18 hrs.
We have identified a face-centered cubic \{111\} structure of a colloidal crystal of nearly close packing.
Nanoparticles isolated with one another have been observed on the surface of gel-immobilized crystal, which can be regarded as gold nanoparticles from their sizes.
\end{abstract}

\newpage
\section*{Introduction}
\label{sec:intro}
When a light is irradiated to a metal, charge density oscillation arises by the electric filed of the light.
Resonance between the electric field caused by this charge density oscillation and the external field is plasmonic resonance.
Plasmon on a surface of a nanosphere, in particular, draws much attention because of strong enhancement of the electric field. 
Such a plasmonic phenonenon is called a localized surface plasmon (LSP).
Enhancement of the optical electric field can be applied not only in the optical information processing but also in chemical and biological sensing.

Photonic crystals are one of branches of the nanophotonics other than the plasmonics.
Periodic variation of dielectric constant with the periodicity of the same magnitude to the optical wavelength exists in a photonic crystal.
Accordingly, photonic band structure takes place.
Lights of a specific wavelength cannot exist in the photonic crystal.
Colloidal spheres arranges in a colloidal crystal regularly with a periodicity of the same magnitude of the optical wavelength.
Colloidal crystals can, therefore, work as photonic crystals.

Recently several authors\cite{Kim2010,Tao2011,Ding2013,Robbinano2013,Xu2013,Belardini2014} fabricated hybrid structures of the LSP and the colloidal photonic crystal.
Hybrid structures with plasmonic nanostructures formed on a photonic crystal are sometimes termed hybrid plasmonic-photonic crystals.
Enhancement of the optical electric field stronger than that due to the LSP alone was reported. 
Au nanoparticles/nanostructure were used to exhibit the LSP resonance by some studies\cite{Robbinano2013,Belardini2014} because of the chemical stability of Au.
Thus, robustness in applications is expected.
On the other hand, Ag nanoparticles/nanostructure were employed in some studies\cite{Kim2010,Tao2011,Xu2013,Ding2013} because of their high performance of the electric field enhancement.
For the photonic crystal material, polystyrene (a regular array of particles or a regular pattern) was selected in some studies\cite{Tao2011,Robbinano2013,Ding2013,Belardini2014}, while poly(methyl methacrylate) was used by Xu et al.\cite{Xu2013} and silica was used by Kim et al.\cite{Kim2010}
An advantage of polystyrene is narrow particle size dispersity.
Polystyrene suspensions with the dispersity less than few percent in the coefficient of variation can be synthesized by a soap-free emulsion polymerization.\cite{Chonde1981}

In particular, Tao, et al.\cite{Tao2011} achieved an enhancement factor up to 260-fold.
We have mentioned above that an intuitive explanation of the enhancement of the optical electric field due to the LSP is based on the directional concentration of the electric field.
Because the light in the photonic band gap is completely reflected by the photonic crystal, when the LSP resonance wavelength and the photonic band gap wavelength coincide with each other, the leakage of the field into the side of the photonic crystal is prohibited.
It will result in the stronger enhancement of the electric field than that by the LSP alone.

A much stronger enhancement is expected by more excellent coincidence between the LSP resonance and the photonic band gap.
Dried colloidal crystals (opals) were used as photonic crystals in previous studies.
In these case, one of ways of tuning photonic band structure is to change the size of colloidal spheres.
The photonic band can, alternatively,  be tuned by altering the volume fraction by Lopez-Garcia et al.\cite{Lopez-Garcia2010} for a hybrid photonic-plasmonic crystal (polystyrene microspheres on an Au substrate).
To accomplish a delicate tuning, gel-immobilized colloidal crystals\cite{Kamenetzky1994,Toyotama2005} are promising.
The photonic band of a gel-immobilized colloidal crystal is tunable due to an external force.\cite{Iwayama2003,Kanai2007}  
An excellent coincidence between the LSP resonance and the photonic band gap is expected due to tuning of the photonic band.
We have tried to replace an opal with a gel-immobilized colloidal crystal.\cite{Kawakami2014}
Reflection spectra in the normal direction were measured for gel-immobilized colloidal crystals with Au nanoparicles deposited on their surface by immersion into an Au nanoparticle aqueous dispersion.
Various immersion time up to 24 hrs (0, 1, 6, 12, and 24 hrs) were tested.
Overall reflectivity for 1 hr sample was lower than that for 0 hr sample.
It increased with immersion time for 6 hr, 12 hr, and 24 hr samples.
Peaks exhibiting photonic band structure were observed for 0 hr and 1 hr samples.
Those peaks disappeared for 6 hr, 12 hr, and 24 hr samples.
Taking into consideration the fact that the enhancement of electric field results from directional concentration, the decrease in reflectivity implies the emission of light to directions other than that for the measurement.
Enhancement of electric field can be expected in this preliminary study.

The purpose in this paper is to observe the nanostructure on a surface of a gel-immobilized colloidal crystal.
Direct detection of a near field of enhanced electric field is not only expensive but also requires specific skills.
Electric field calculation is helpful to complement.
To this end, one should know the structure in detail. 
Because in a preliminary study\cite{Kawakami2014} samples with immersion time longer than 6 hrs exhibited feature-less reflectivity spectrum, we prepared samples for the immersion time up to 2 hrs.
We speculated that for such a long immersion time an Au thin film formed.
In other words, it is a strategy to start with simple structures formed for a short immersion time and then to proceed complicated structure formed for a long immersion time.
However, if the number density of nanoparticles on the surface is too low, cases can be arisen that no particles are seen in the observe area.
We report in this paper results of atomic force microscope (AFM) observation for a 2 hr sample.

\section*{Experimental}
\subsection*{Materials}
\label{sec:mater}
Polystyrene colloidal suspension synthesized by a soap-free emulsion polymerization\cite{Chonde1981} were used to make colloidal crystals.
Styrene and divinilbenzene were purchased from Tokyo Kasei (S0095 and D0958, respectively), 4-styrensulfinic acid, sodium salt hydrate from Aldrich (328596-100G) and potassium presulfate from Sigma-Aldrich (216224-100G).
A scanning electron microscope (SEM) image of the colloidal spheres is shown in Fig.~\ref{fig:PSt-SEM}.
A field-emission type SEM (Hitachi S4700) was used.
Before taking SEM images dried colloid dispersion adhered on a carbon tape was coated with Pt-Pd by spattering.
The thickness of the coat was about 5 nm.
We analyzed the SEM image by a software.
$190\pm10\mbox{ [nm]}$ was obtained for the diameter of the colloidal sphere. 
The density of the colloidal suspension measured was $7\pm1$ [\%] in volume fraction.

The colloidal suspension was stocked with addition of ion exchange resins (BidRad AG501-X8).
Iridescence appeared in the stocked suspension.
We shook this suspension with ion exchange resins for weeks.
Gelation of the colloidal crystals was accomplished essentially according to Toyotama et al.\cite{Toyotama2005}
The gel regent ($N$-(Hydroxy methyl)acrylamide), the initiator (2,2'-azobis[2-methyl-$N$-(2-hydroxyethyl)propionamide]), and the cross-linker ($N$,$N$'-methylen-bis-acrylamide) were purchased from Wako Chemical (081-01535, 929-11852, and 929-41412, respectively).
After mixing aqueous solutions (abbreviated as aqs. in Table~\ref{tab:gel}) of those chemicals to the colloidal suspension (amount of chemicals are listed in Table~\ref{tab:gel}), babbling with an argon gas to remove gases which prevent polymerization of the gel monomers such as oxygen and carbon dioxide was done for three minutes prior to the gelation.
We note that in a preliminary study\cite{Kawakami2014} the gel regent was purchased from Sigma-Aldrich and a nitrogen gas was used for bubbling.
Milli-Q water (MQW) with electric conductivity higher than 18.2M $\Omega\cdot$cm was used.
After injecting the mixed suspensions into a quartz cell with thickness 1 mm, we gellated the samples by ultra violet irradiation for thirty minutes.
It is noted that efforts was not made to improve the photonic property of the colloidal crystal because the purpose of this paper is to observe the surface structure (in particular, interrelation between Au nanoparticles and colloidal spheres).
Among many factors a relatively wide size dispersity of colloidal spheres is pointed out.

The gel-immobilized colloidal crystals were taken out from the thin quartz cell.
And then the gel-immobilized colloidal crystals were immersed into an aqueous dispersion of Au nanoparticles.
Immersion times of 0, 0.5, 1.0, 1.5, and 2.0 hrs were tested.
The aqueous dispersion of Au nanoparticles (40nm in diameter) was synthesized according to the recipe in Morandi, et al.\cite{
Morandi2007}
The number density of Au nanoparticles in this dispersion was less than a few thousandths of percent in volume fraction.

\subsection*{Measurement}
\label{sec:measure}
As-prepared gel-immobilized colloidal crystals are wet.
Therefore, observation of surfaces of such samples by contact-mode AFM accompanies difficulties.
Electron microscope observations are, of course, impossible.
In this study, AMF observations were done for dried samples with a contact mode.
The AFM used for the observations was equivalent to Shimadzu SPM-8000FM.

\section*{Results}
\subsection*{Drying Samples}
\label{sec:dry}
As mentioned above, for the AFM observations we prepared dried samples.
We put the samples on a sample stage and then dried them in a desiccator for 18 hrs.
Photographs of 0 hr and 2 hr samples just after picking-up from a water and after drying are shown in Fig.~\ref{fig:drying}.
We see shrink of both samples in the parallel direction to the stage by drying.
The linear shrink ratio is fifty and several percent.
As far as Fig.~\ref{fig:drying} shows, the shrink is isotropic in the parallel direction.
It does not mean a three-dimensional isotropic shrink.
As mentioned already the thickness of the gel-immobilized colloidal crystals was 1 mm.
Thus, accuracy in measurement of shrink in the normal direction was limited.

\subsection*{AFM Results}
\label{sec:AFM}
We observed two different positions of the surface of a 2 hr sample.
One set of results of the AFM observations is shown in Fig.~\ref{fig:PSt-AFM}.
In a topographic image [Fig.~\ref{fig:PSt-AFM}~(a)], many hexagons made of six particles at verteces and one central particle can be recognized.
This is an indication of a face-centered cubic (fcc) \{111\} structure.
A differential image, where height difference is emphasized, is shown Fig.~\ref{fig:PSt-AFM}~(b).
Why is the sizes of spots roughly estimated in Fig.~\ref{fig:PSt-AFM}~(a) smaller than the diameter of colloidal spheres?
Naive interpretation is that top portions of colloidal spheres, not entire ones, may rise above the gel matrix.
Height profiles along lines AB and CD in Fig.~\ref{fig:PSt-AFM}~(a) [Fig.~\ref{fig:PSt-AFM}~(c) and (d), respectively] give a support to this conjecture.
The height difference between a peak and its adjacent bottom can be understood as more than 90\% portion of a colloidal sphere being embedded in the gel matrix.
Another interpretation is possible, but it does not alter the essence.
The surface might be covered by a thin gel film.
The thickness was, however, so small that the surface wandered reflecting the height variation of the fcc \{111\} face. 
The latter interpretation is more comprehensive.
Thickness of gel coat on the sphere after shrink may not be a linear function of the thickness of gel before shrinking under vertical compressive stress.
As a possibility, the thickness after shrink can become zero if the thickness before shrinking is less than certain value.
We will discuss more in Discussion section.

We have additional information in Fig.~\ref{fig:PSt-AFM}~(b).
Tiny spots are seen, which may be images of the Au nanoparticles.
To confirm this speculation we must exclude a possibility that these are just noises.
If no Au nanoparticles are observed, the surface number density of Au nanoparticles is so low that no Au particles exist in the present observation area.
Even if these spots are of Au nanoparticle, the particle number density is enough low so that complicated clusters does not form.
This discussion is supported by the high dilution of the present Au nanoparticle dispersion.

An AFM result of another position on the surface of the 2 hr sample are shown in Fig.~\ref{fig:Au-AFM}.
At first, we consider on large spots in Fig.~\ref{fig:Au-AFM} (a).
In a topographic image [Fig.~\ref{fig:Au-AFM}~(a)], the particle number density is low as compared with Fig.~\ref{fig:PSt-AFM}~(a).
And, accordingly, unlike Fig.~\ref{fig:PSt-AFM}~(a), the particles arranges irregularly.
With respect to the large spots, interpretation of a differential image [Fig.~\ref{fig:Au-AFM}~(b)] is entirely the same as that for Fig.~\ref{fig:PSt-AFM}~(b).

Now, we turn to tiny spots in Fig.~\ref{fig:Au-AFM}~(a).
Prior to looking into a magnification [Fig.~\ref{fig:Au-AFM}~(c)], let us look at Fig.~\ref{fig:Au-AFM}~(b).
In this figure one can see particles at the corresponding position of the tiny spots.
However, the situation is the same as that for Fig.~\ref{fig:PSt-AFM}~(b); the possibility of noises are not entirely excluded.
Fig.~\ref{fig:Au-AFM}~(c) is a magnification of a tiny spot pointed by an arrow in Fig.~\ref{fig:Au-AFM}~(a).
One can estimate the length of long axis as c.a.~40nm and that of short axis c.a.~30nm.
We infer that the tiny spots are images of the Au nanoparticles.
It should be noted that most of the Au nanoparticles were isolated with one another.
At last, we look at the height profile [Fig.~\ref{fig:PSt-AFM} (d)].
The difference between maximum height, say $Z(X_m)$, and the height at $X_m\pm 20$ nm is one and several tenths of nanometers.
Here, $Z(X)$ is the hight at position $X$, and $X_m$ denotes a position giving the maximum height.
Even if one employs the value of $Z(X_m)$ itself (detection of particle edges with an AFM is inaccurate), it is a few nanometers.
These values are much smaller than the diameter of the Au nanoparticle.
Of course, interpretation given for the colloidal spheres is possible for the Au nanoparticles, too. 
Mechanisms with regard to shrink of gel are, however, different from each other.
We will discuss more in Discussion section.

Analyses of height profiles across anther nine spots in Fig.~\ref{fig:Au-AFM}~(a) similarly to Fig.~\ref{fig:Au-AFM}~(c) and obtained results similar to Fig.~\ref{fig:Au-AFM}~(d).
The height profiles depend on ratio of embedded portion.
That is, the profile with wide width has the lower peak height.
It means that a sphere is covered by a thicker gel film for the lower height profile.
Since the relation between the width and height is unimodal, a monodisperse size distribution is inferred.
Such monodispersity is not expected in polymerization of gel regents without delicate control.
In addition, such spots were not observed for samples of 0 hr immersion; AFM images like Figs.~\ref{fig:PSt-AFM} and \ref{fig:Au-AFM} without such spots were obtained such as in Fig.~A2.
In this way, it is has been shown that the images c.a.~40 nm were certainly related to Au nanoparticles. 
Nevertheless, possibility of gel-origination of such objects has not been ruled put completely.
That is, gel-originated objects might be formed induced by Au nanoparticles.
If one successfully performed composition analysis, an ultimate evidence is obtained.
We tried to an energy dispersive X-ray spectroscopy (EDX).
Unfortunately, the Au peak in an EDX spectrum even for a dried 40 nm Au nanoparticle dispersion was very faint.
That is, without notifying the targeted composition of Au it was difficult to specify the composition from the EDX spectra alone.
The Au peaks in the EDX spectra were undetectably faint for 2 hr immersion samples.
We emphasize here that the AFM images of Au nanoparticles in dried Au nanoparticle dispersion were indistinguishably similar to 40 nm objects in Fig.~\ref{fig:Au-AFM}.
Because AFM peoples know that detection on edge is not accurate due to the extent of tip, they must regard the degree of this similarity is within the coincidence.
A detail of this AFM analysis is presented in Appendix A.

\section*{Discussion}
\subsection*{Interparticle Spacing in fcc \{111\}}
\label{sec:space}
We evaluate interparticle spacing of colloidal sphere.
Let be the volume fraction of colloidal spheres in the stock colloidal suspension $\phi_{col}$ = 7 $\pm$ 1 [\%].
The total volume of colloidal spheres containing the suspension of volume $V_{col}$ = 700 $\mbox{$\mu$l}$ is given by $\phi_{col} V_{col}$.
The total volume of mixed liquid of colloidal suspension and gel regent (volume $V_{gel}$ = 150 $\mbox{$\mu$l}$) with initiator and cross-linker solutions (volumes $V_{ini}$ = 100 $\mbox{$\mu$l}$ and $V_{lin}$ = 50 $\mbox{$\mu$l}$, respectively) is approximately given by $V_{tot}$ = $V_{col}$ + $V_{gel}$ + $V_{ini}$ + $V_{lin}$.
Volume fraction of the gel-immobilized colloidal crystal is given by $\phi_{gic}$ = $\phi_{col} V_{col}/V_{tot}$ and evaluated as $\phi_{gic}$ = 5 $\pm$ 1 [\%].

Let us evaluate the lattice constant and interparticle spacing assuming that the present gel-immobilized colloidal crystal is an fcc crystal.
This assumption is valid for concentrated colloidal suspensions.
For an fcc crystal of spheres with diameter $\sigma$ the volume fraction is given by $\phi$ = $2\pi\sigma^3/3a^3$, where $a$ is the lattice constant.
Thus, the lattice constant is calculated by $a$ = $(2\pi/3\phi)^{1/3} \sigma$.
Letting $\sigma$ be 190 $\pm$ 10 [nm] and substituting $\phi$ = 0.05 $\pm$ 0.01, we have $a$ = 660 $\pm$ 40 [nm].
The interparticle spacing in an fcc \{111\} plane is related to $a$ as $l_{111}$ = $a/\sqrt{2}$.
Accodingly, we evaluate $l_{111}$ = 470 $\pm$ 30 [nm].
Letting the shrink ratio be $0.55\pm 0.05$, we obtain the interparticle spacing for the dried gel-immobilized crystal as $l_{111}^{dry}$ = 260 $\pm$ 20 [nm].

Let us estimate the interparticle spacing from the height profile [Fig.~\ref{fig:PSt-AFM}~(c)].
The horizontal positions of two adjacent particles on the slightly tilted line are marked by crosses in Fig.~\ref{fig:PSt-AFM}~(a) and corresponding positions are indicated by vertical lines in Fig.~\ref{fig:PSt-AFM} (c).
The separation of these lines is c.a.~260 nm.
A similar analysis can be done for Fig.~\ref{fig:PSt-AFM}~(d), resulting in the same estimation.
Surprisingly, this value does just coincide to the value of $l_{111}^{dry}$ calculated just above.
Looking at the horizontal line in Fig.~\ref{fig:PSt-AFM}~(a), height profiles were measured along lattice lines.
However, the values of peaks are different from one another.
Nevertheless, the magnitude of hight difference is negligibly small compared to the size of the spheres.
Conclusively, an AFM result [Fig.~\ref{fig:PSt-AFM}] can be understood as an fcc \{111\} face of the gel-immobilized colloidal crystal of nearly close packing.

\subsection*{Surface Color}
The surface color of samples in Fig.~\ref{fig:drying} (a) is faint.
We discuss here on the surface color of dried samples [Fig.~\ref{fig:drying} (b)].

The surface color 2 hr sample can be understood as the complementary color of adsorption due to the LSP of 40 nm Au nanoparticles (c.a.~540 nm).
Shrink of sample by drying concentrated the faint color of wet sample [right one in Fig.\ref{fig:drying}~(a)] into the deep one.
Other interpretations are possible, but this understanding is quite reasonable.
In previous work \cite{Kawakami2014} we have already concluded that Au nanoparticles adhered on the gel surface by immersion.
We infer the Au nanoparticles existed on the surface of 2 hr sample.

Surface color of 0 hr sample can be interpreted from the phtonic band of the colloidal crystal.
It is concluded that the dried colloidal crystal was almost a close-packed one.
An AFM result [Fig.~A2] support this property also for 0 hr sample.
The lattice constant of close-packed fcc crystal of $\sigma$ = 190 nm spheres is $a$ = $\sqrt{2}\sigma$ $\sim$ 270 nm.
The interlayer spacing between fcc \{111\} is $d_{111}$ = $a/\sqrt{3}$ $\sim$ 160 nm and the corresponding Bragg wavelength evaluated from $m\lambda$ = $2n_{av} d \sin\theta$
with $m$ = 1 and $\theta$ = $90^{\circ}$ is $\lambda_{111}$ $\sim$ 450 nm.
This interprets the blue color.
The same evaluation for the (002) reflection is $\lambda_{002}$ $\sim$ 390 nm, interpreting the violet color.
It is more reasonable that the (001) reflection of $\lambda_{001}$ $\sim$ 770 nm (rad) is mixed.
In those evaluation the effective refractive index has been approximated by arithmetic average $n_{av}$ $\cong$ $\phi_{cp} n_{PSt}$ + $(1-\phi_{cp})n_{air}$ $\sim$ 1.44 with $\phi_{cp}$ = $\sqrt{2}\pi/6$ begin the close-pack volume fraction and $n_{PSt}$ and $n_{air}$ the refractive indices of respective materials.
Mixing of (001)-oriented domains in \{111\}-oriented crystals is suggested.
We have already mentioned about the insufficient crystallinity in Materials subsection.
It is noted that because the densification is the driving force of colloidal crystallization, drying improves the crystallinity locally.

We have mentioned in Materials subsection that efforts to improve the photonic property.
Confirmation of both the plasmonic adsorption and the photonic band gaps through reflection spectrum was difficult for such samples.
In particular, adsorption measurement itsself was difficult because of the sample thickness of 1 mm.

\subsection*{Shrink of Gel}
\label{sec:shrink}
We have already discussed on a gel film covering a colloidal sphere.
Here, discussion regarding shrink of gel by drying on an effect of hight profiles across both the colloidal spheres and the Au nanoparticles.

At first, we consider a colloidal sphere embedded in a gel matrix such as illustrated in Fig.~\ref{fig:col-gel}.
A gel-immobilized colloidal crystal is made from a mixed liquid of a colloidal suspension and gel materials.
Hence, most of colloidal spheres locate inside the gel.
So, one can have a schema like in the left one of Fig.~\ref{fig:col-gel}.
A portion of gel on the top of the colloidal sphere is incurred from a vertical compressive stress because that portion is sandwiched by the gel surface and the top of sphere.
We note that although nothing is pushing the the gel surface downward, the surface tension play a role.
The part far from the colloidal sphere may be under a negligible stress.
In such an anisotropic and non-uniform stress condition, the gel shrinks anisotropically and non-uniformly.
Thus, the portion of gel on the top of colloidal sphere becomes thiner like in the middle schema of Fig.~\ref{fig:col-gel}.
Eventually, the top of colloidal sphere may rise above the surface like in the right schema of Fig.~\ref{fig:col-gel}.
A situation like in the middle schema of Fig.~\ref{fig:col-gel}, where the top of colloidal sphere is covered by a gel film, can occur after enough drying in a case that the portion of gel on the top of colloidal sphere is enough large.
One can understand wandering of a gel surface based on the stress from colloidal spheres.

Next, we consider an Au nanoparticle on a gel surface such as illustrated in Fig.~\ref{fig:Au-gel}.
Unlike the case of the colloidal sphere [Fig.~\ref{fig:col-gel}], initially Au nanoparticles locate outside the gel matrix because a gel-immobilized colloidal crystal is immersed into an Au nanoparticle dispersion.
Let us start with a situation like in the left schema of Fig.~\ref{fig:Au-gel}, where an Au nanoparticle is attached on a gel surface.
The Au nanoparticle gives a downward compression.
Under such a stress state the gel shrinks anisotropically and non-uniformly.
A more compressed part sinks more than less compressed part.
It will result in a situation like in the right schema of Fig.~\ref{fig:Au-gel}.
This scenario explains the height profile with the amplitude smaller than the diameter of the Au nanoparticle.
As discussed at the last of Result the particles exhibiting low peak height in the height profile are suggested to be covered by a gel film.

\section*{Concluding Remarks}
We prepared gel-immobilized colloidal crystals and immersed them into a gold nanoparticle dispersion.
A surface of a sample has been observed by an AFM after drying.
The obtained results are summarized as follows.
\begin{itemize}
\item An fcc \{111\} structure formed by colloidal spheres in surface region of the sample was observed. 
\item Gold nanoparticles isolated with one another were identified on the sample surface .
\item We gave an interpretations for spatial oscillation of hight profiles with amplitudes much less than the size of colloidal sphere based on a non-uniform anisotropic stress provided by the particles.
\item Also, an interpretation were given for the same phenomena for Au nanoparticles on the same basis.
\end{itemize}

The secondly itemized one is a piece of key information toward proceeding studies.
The LSP mode of single metal nanoparticle is understood theoretically and a lot of electric field calculations have been performed.
A hybrid mode may be understood as an extension of the LSP of single particle.
We can have a strategy for electric field calculations on the hybrid structure base on the second result.

Let us give other remarks for future research regarding observation of the nanostructures.
As discussed in Discussion section surface morphology is suffered from shrink of a gel in drying process.
We have a plan to observe wet surfaces by a non-contact mode AFM.
Moreover, observation of cross sections of samples by the non-contact AFM will be a next one.
Hydrogels contain many amount of water such as 90\% or more in weight fraction.
It means that a metal nanoparticle can move in pores in a gel network geometrically.
As a possibility, formation of three-dimensional nanostructure is suggested.
Observation of samples as wet may resolve issues regarding the covering by a gel film for size determination from height profiles across nanoparticles.

Besides the plan of observation of wet surface, some remarks are given at the end of paper.
It was annoying that for both 0 hr and 2 hr immersion samples there were nanometer-size images which were completely different from those in Fig.~\ref{fig:Au-AFM} and Fig.~A1.
While for 18 hr dried samples such image could not be found frequently, for samples with drying of months long we encountered such image more frequently.
It means that drying gels induces surface nanostructure other than Au nanoparticle attachment.
The amount of such structure increases with the drying time.
No such structure can, therefore, be expected for as-prepared samples.

\section*{Acknowledgment}
It is acknowledged that Dr. T. Sawada (NIMS) contributed to this work equally to the authors while he was at his former position.
Authors thank Dr. Y. Suzuki (Tokushima Univ.) for helpful discussions.
This work has been supported by a research grant from The Mazda Foundation in 2013.

It is acknowledged that one of referees pointed another potential of Au nanoparticles.
Concerning the fact that this paper has not ruled out completely the possibility that the objects c.a.~40 nm seen in AFM images were made from gel material, gel or organic particles or structure c.a.~40 nm with tiny size dispersity can form under presence of Au nanoparticles like a catalytic effect.
A field of new technique may be opened.

\appendix

\section*{Appendix 1: AFM result of Au nanoparticles}
An AFM result for dried Au nanoparticle dispersion is shown in Fig.~A1.
An drop of Au nanoparticle dispersion was put on a silicon substrate.
After drying AFM observation was performed.
A number of drying times were tested.
Under slow evaporation of dispersion medium nanoparticles form colloidal crystals or aggregations.
In Fig.~A1 aggregations are seen although we tried to re-disperse the aggregated Au nanoparticles by adding a tiny amount of water and then drying again.
The appearance is quite similar to that in Fig.~\ref{fig:Au-AFM}.
Since in this sample the substrate is flat, the peak height in Fig.~A1~(c) does coincide to the particle diameter of c.a.~40 nm.
On the other hand, the widths estimated along $X$ axis in the height profiles [Fig.~\ref{fig:Au-AFM}~(d) and Fig.~A1~(c)] do not exhibit excellent coincidence.
It is commonly known that one of shortcomings of AFM is not good at detecting edges.
Also it is noted that Fig~\ref{fig:Au-AFM} (c) is a magnification of a wide area image [Fig.~\ref{fig:Au-AFM} (a)].
The resolutions are, therefore, different for Fig.~\ref{fig:Au-AFM}~(c) and Fig.~A1~(a).
It is concluded that one cannot distinguish the c.a.~40 nm image in Fig.~\ref{fig:Au-AFM}~(c) from the images of a 40 nm Au nanoparticle in a framework of AFM.

\section*{Appendix 2: AFM result of 0 h immersion sample}
An AFM result for 0 hr immersion sample is shown in Fig.~A2.
AFM observations were performed after drying of 18 hrs as so for 2 hr sample.
The analysis supports that the samples were almost close-packed after drying.
It allows interpretation of surface color given in Surface Color subsection.
It is noted that the image of Fig.~A2 includes an artifact caused by the scanning direction of $X$.
Lines in this  direction are seen on the particle images.
There is a line on left particle under the line in Fig.~A2~(a) along which the height profile was analyzed.
A spike in height profile [Fig.~A2~(c)] is attributed to this artifact.

It is noticeable that there are no c.a.~40 nm objects similar to those in Figs.~\ref{fig:PSt-AFM} and \ref{fig:Au-AFM} and Fig.~A1.
As mentioned at the end of Concluding Remarks section, there were objects whose appearance was completely different from that of Au nanoparticles.
We repeat here that c.a.~40 nm objects in Figs.~\ref{fig:PSt-AFM} and \ref{fig:Au-AFM} are certainly Au-nanoparticle related.

\newpage
\section*{Figure legends}

\begin{figure}[t]
\begin{center}
\includegraphics[width=0.75\hsize]{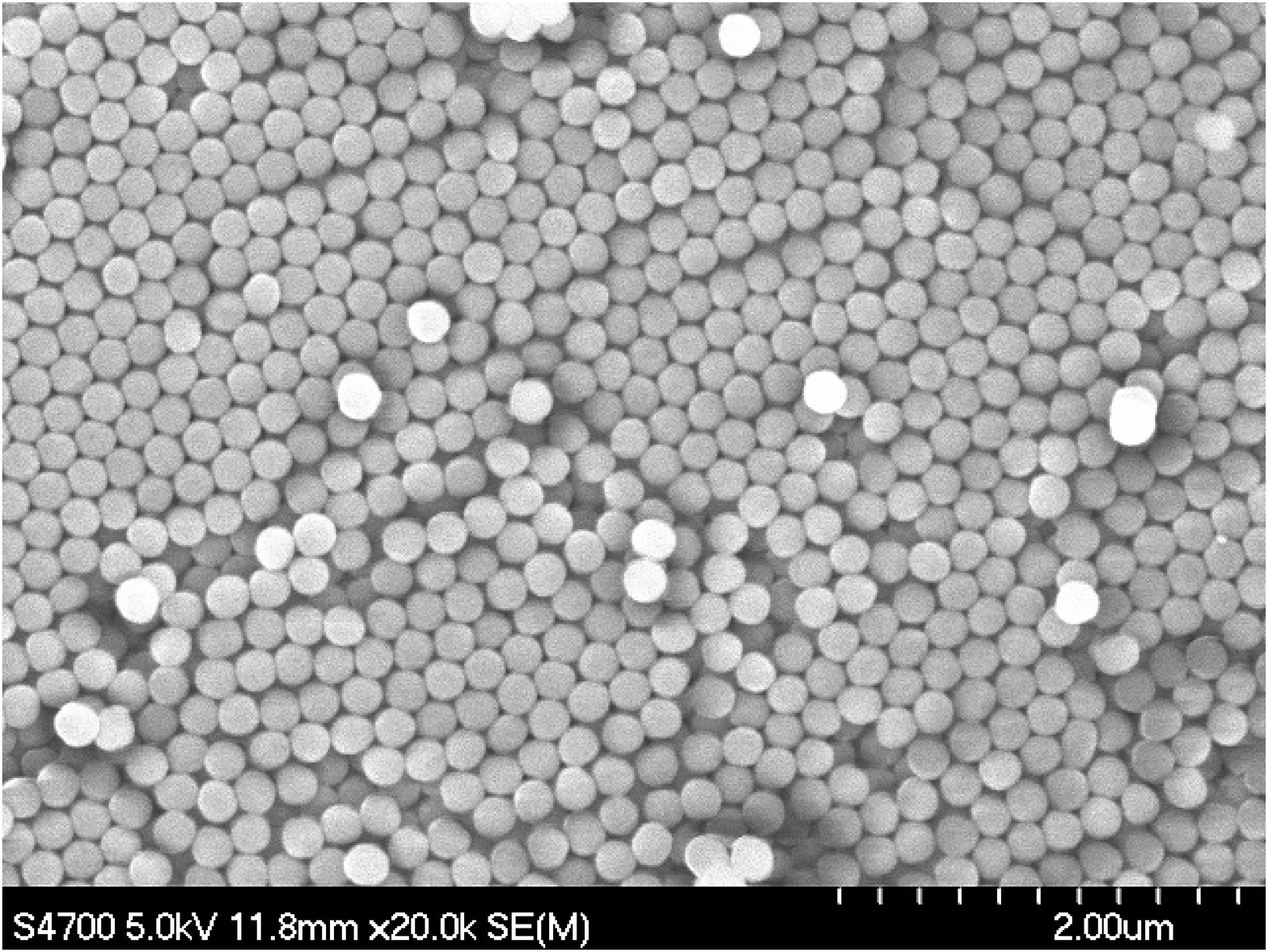}
\caption{\label{fig:PSt-SEM}}
\end{center}
\end{figure}

\noindent
Figure \ref{fig:PSt-SEM}:
An SEM image of colloidal spheres used in this study. \\ 

\begin{figure}[t]
\begin{center}
\begin{tabular}{cc}
\begin{minipage}[t]{0.5\hsize}
\begin{flushright}
\includegraphics[width=\hsize]{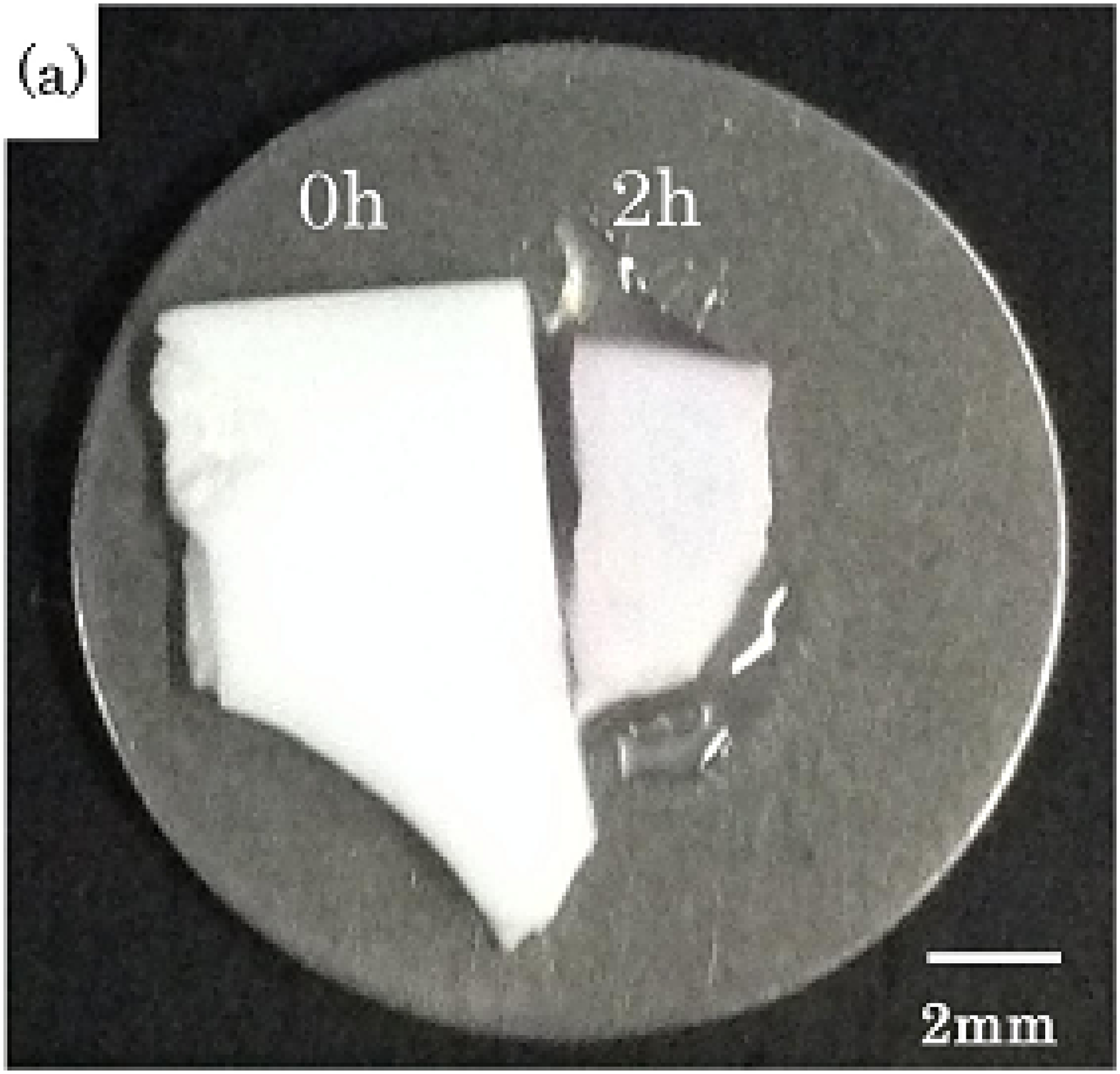}
\end{flushright}
\end{minipage} &
\begin{minipage}[t]{0.5\hsize}
\begin{flushleft}
\includegraphics[width=\hsize]{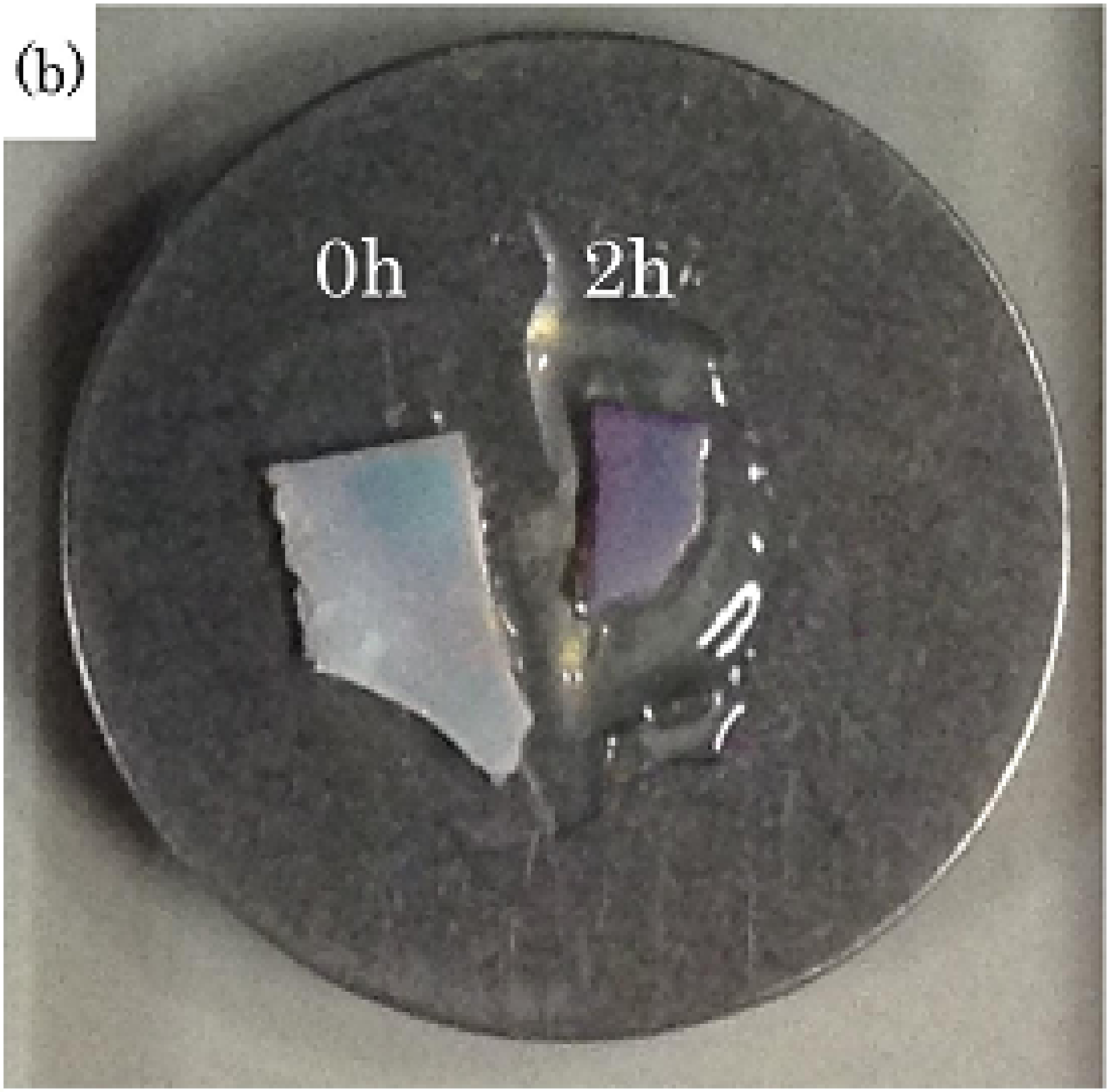}
\end{flushleft}
\end{minipage}
\end{tabular}
\caption{\label{fig:drying}}
\end{center}
\end{figure}

\noindent
Figure \ref{fig:drying}: Outer shapes of samples: (a) before and (b) after drying.
Immersion times are indicated in the photographs.
Samples shrunk by fifty and several percent by drying. \\

\begin{figure}[t]
\begin{center}
\includegraphics[width=0.75\hsize]{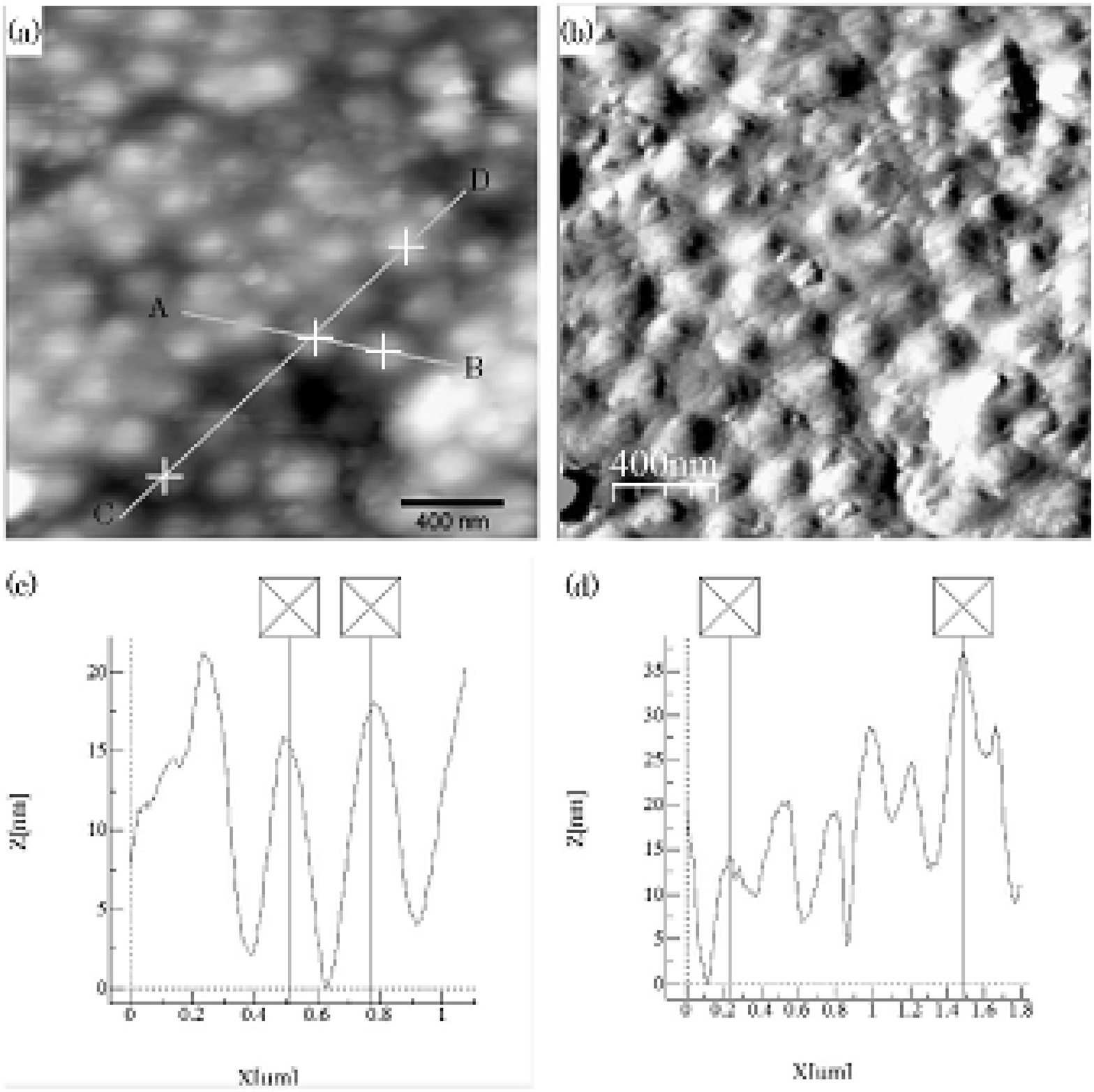}
\caption{\label{fig:PSt-AFM}}
\end{center}
\end{figure}

\noindent
Figure \ref{fig:PSt-AFM}:
Results of one of an AFM observation: (a) a topographic image, (b) a differential image, (c) a height profile along line AB in (a), and (d) that along line CD in (a).
In (a), hexagons formed by six particles at verteces and one central particle are seen.
This is an indication of an fcc \{111\} plane.
In (b), top portions of spheres may be observed. \\

\begin{figure}[t]
\begin{center}
\includegraphics[width=0.75\hsize]{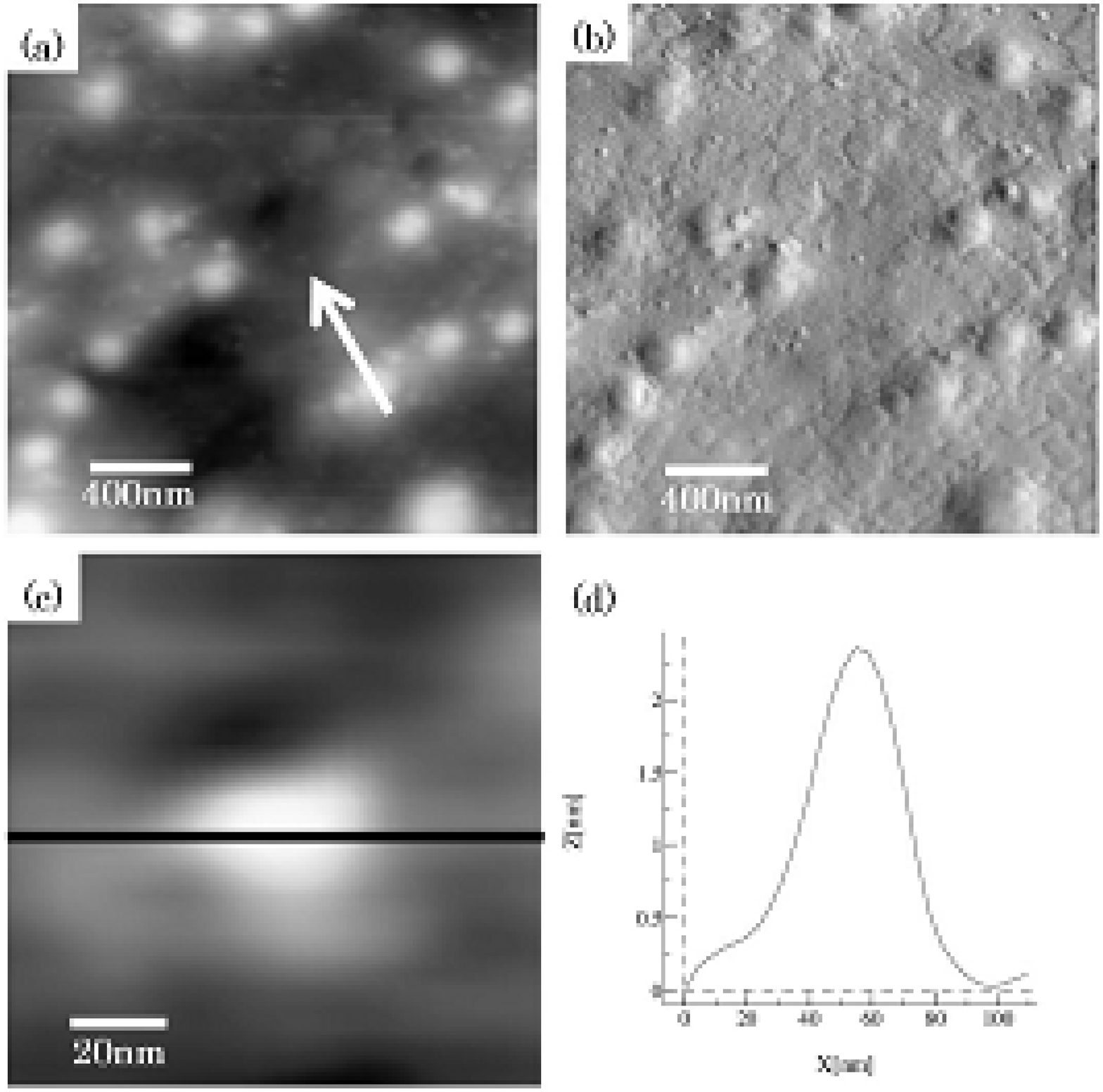}
\caption{\label{fig:Au-AFM}}
\end{center}
\end{figure}

\noindent
Figure \ref{fig:Au-AFM}:
Results of an AFM observation of a position different from that in Fig.~\ref{fig:PSt-AFM}: (a) a topographic image, (b) a differential image, (c) a magnification of tiny spot pointed by an arrow in (a), and (d) a height profile along the line in (c).
In (a) and (b), colloidal particles arranged irregularly.
From the size of spot in (c), the tiny spots are likely images of the Au nanoparticle.
The height profile (d) indicates a sink of an Au nanoparticle below the surface of the gel matrix. \\

\begin{figure}[t]
\vspace*{5cm}
\begin{center}
\includegraphics[width=0.75\hsize]{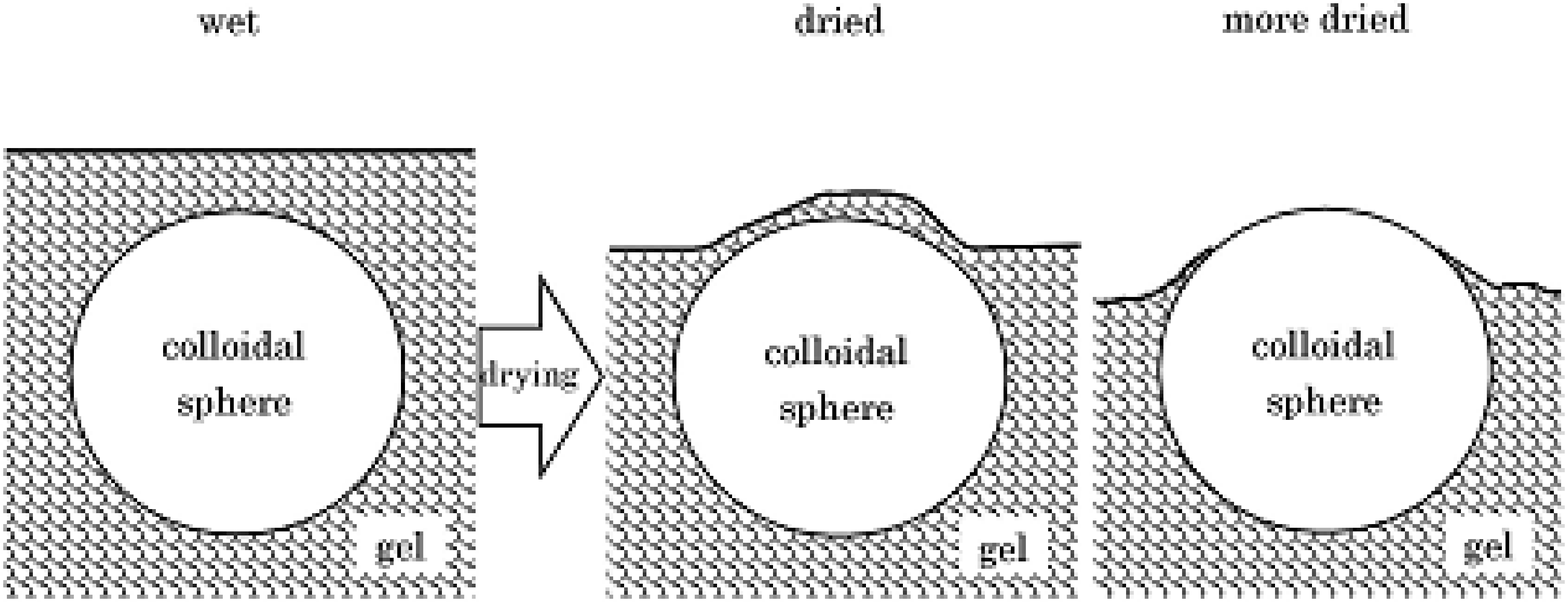}
\caption{\label{fig:col-gel}}
\end{center}
\end{figure}

\noindent
Figure \ref{fig:col-gel}:
An illustration of a colloidal sphere in a gel in drying process.
A portion of gel on the top of colloidal sphere is incurred from a vertical compressive stress.
In process of drying under such a stress sate, the gel shrinks anisotropically and non-uniformly.
In a case the colloidal sphere will still be covered by gel, in other case the top of colloidal sphere will rise above the surface. \\

\begin{figure}[t]
\begin{center}
\includegraphics[width=0.9\hsize]{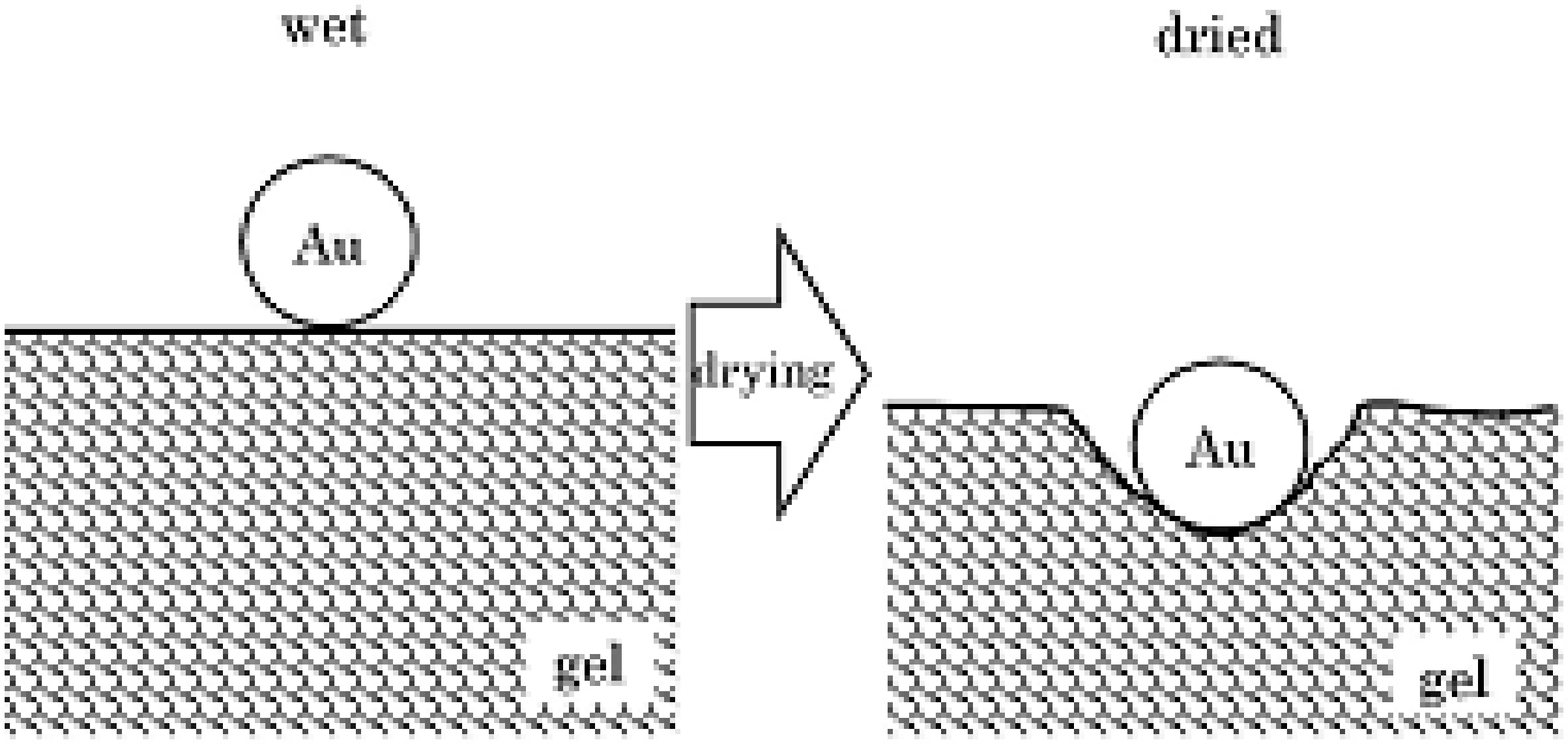}
\caption{\label{fig:Au-gel}}
\end{center}
\end{figure}

\noindent
Figure \ref{fig:Au-gel}:
An illustration of an Au nanoparticle on a gel in drying process.
A portion of gel beneath the bottom of nanoparticle is incurred from a vertical compressive stress.
In process of drying under such a stress sate, the gel shrinks anisotropically and non-uniformly. \\

\noindent
Figure A1:
Results of an AFM observation of dried Au nanoparticle dispersion: (a) a topographic image, (b) a differential image, (c) a height profile along a line in (a). \\

\noindent
Figure A2:
Results of an AFM observation of 0 h immersion sample: (a) a topographic image, (b) a differential image, (c) a height profile along a line in (a).

\newpage
\begin{table}[t]
 \begin{center}
\caption{Amount of chemicals for gel-immobilized colloidal crystals.}
\label{tab:gel}
  \begin{tabular}{c||c|c} \hline
    materials & states & amounts \\ \hline \hline
    colloidal suspension &  stock suspension & 700 $\mu$l \\ \hline
    gel regent & aq. (3.0 g with 2 ml MQW) & 150 $\mu$l \\ \hline
    initiator & aq. (0.02 g with 5 ml MQW) & 100 $\mu$l \\ \hline
    cross-linker & aq. (0.077 g with 5 ml MQW) & 50 $\mu$l \\ \hline
  \end{tabular}
 \end{center}
\end{table}

\newpage
\vspace*{\fill}

\newpage
\vspace*{\fill}

\vspace*{\fill}

\vspace*{\fill}

\vspace*{\fill}

\vspace*{\fill}

\begin{figure}[t]
\begin{center}
\includegraphics[width=0.75\hsize]{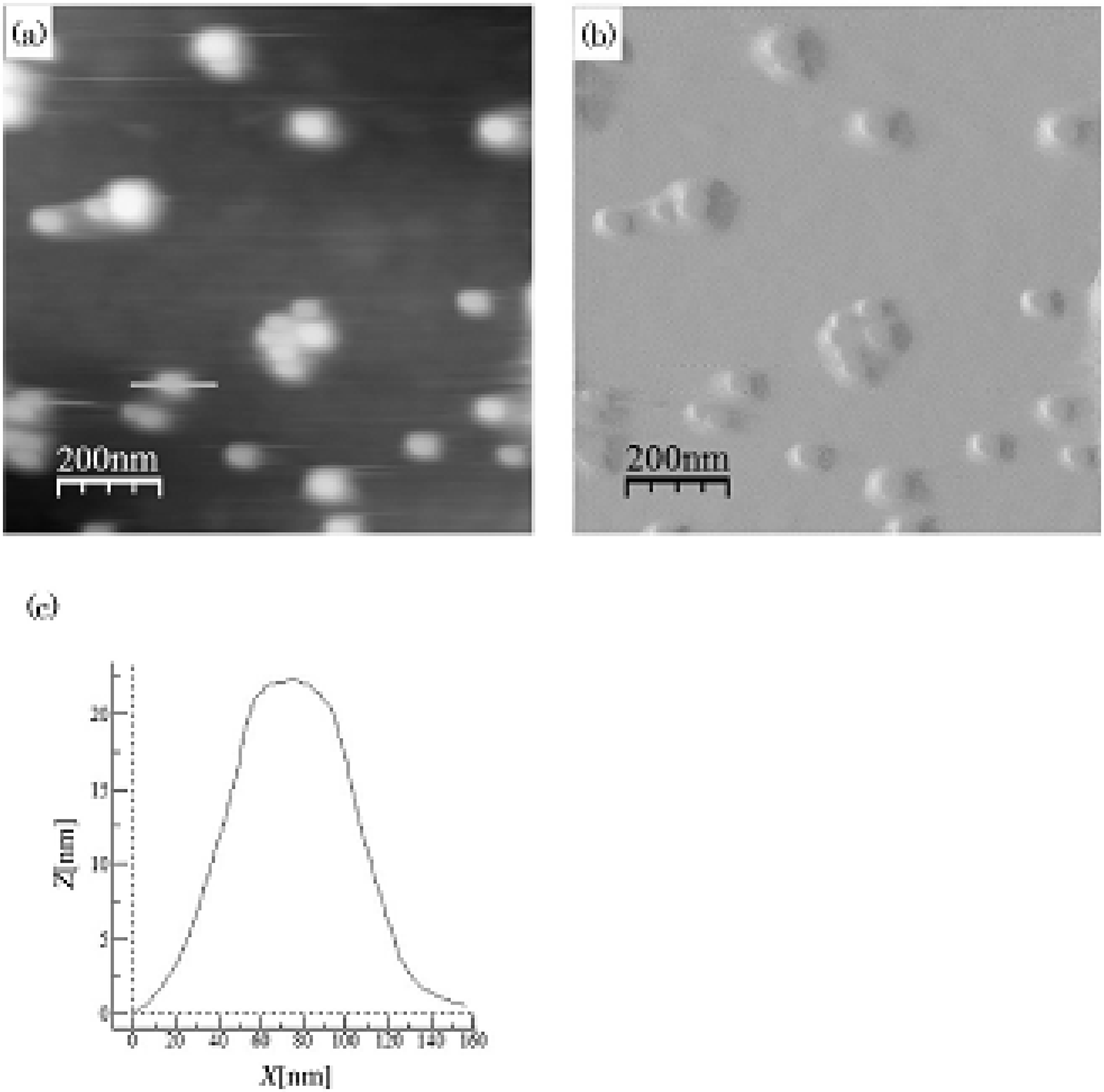} \\
Figire A1
\end{center}
\end{figure}
\vspace*{\fill}

\newpage
\begin{figure}[t]
\begin{center}
\includegraphics[width=0.75\hsize]{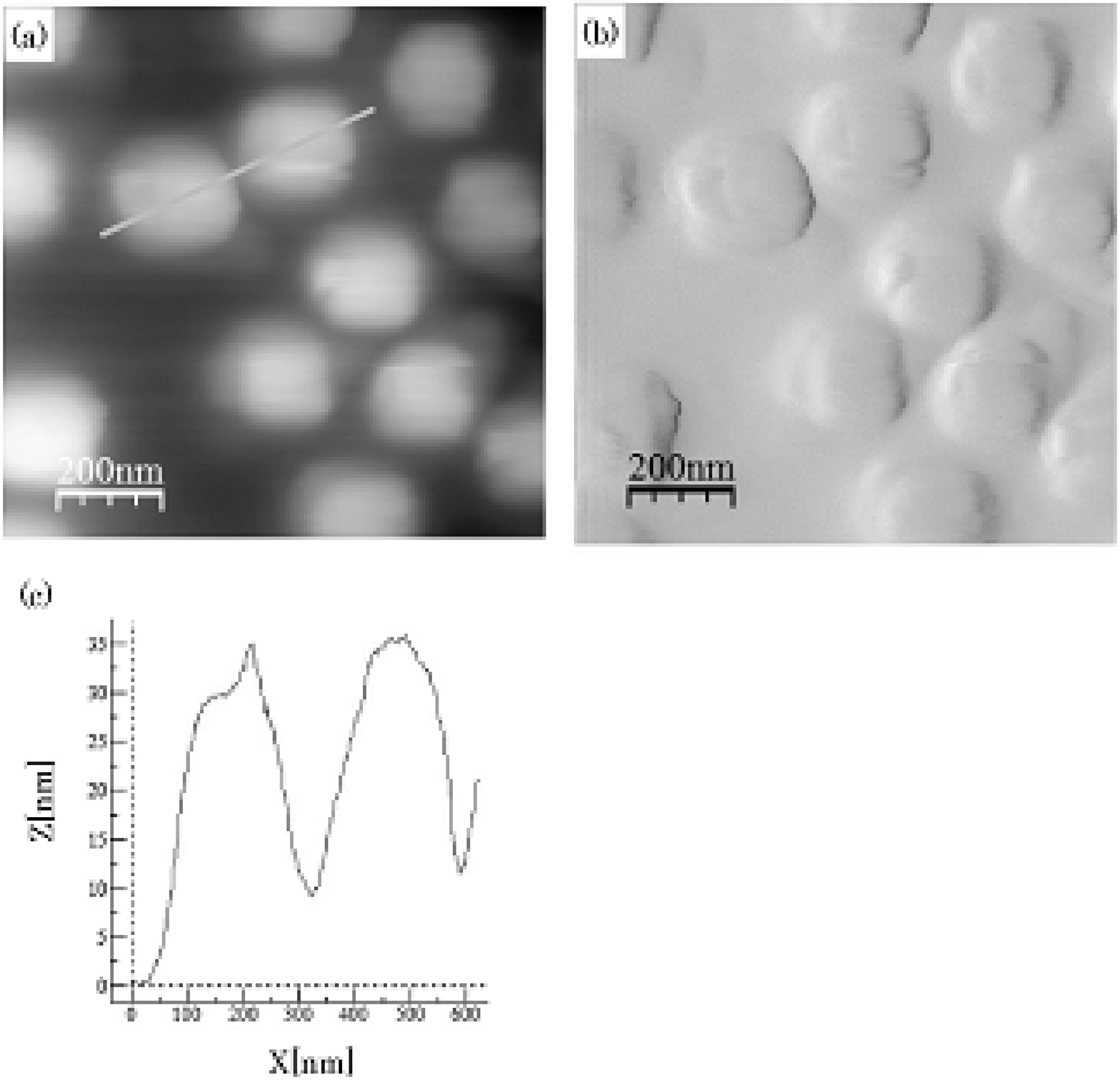} \\
Figure A2
\end{center}
\end{figure}
\vspace*{\fill}

\end{document}